\documentclass[pra,showpacs,twocolumn,superscriptaddress,floatfix]{revtex4-1}

\usepackage[dvipsnames]{color}
\usepackage{graphicx}
\usepackage{bm}

\newcommand{\be}{\begin{equation}}
\newcommand{\ee}{\end{equation}}
\newcommand{\bea}{\begin{eqnarray}}
\newcommand{\eea}{\end{eqnarray}}

\newcommand{\mbf}[1]{\mathbf{#1}}
\newcommand{\abs}[1]{\left|#1\right|}

\begin{document}
\title{Metric-space {approach for distinguishing} quantum phase transitions in spin-imbalanced systems}
\author{T. de Picoli}
\affiliation{S\~{a}o Carlos Institute of Physics, University of S\~{a}o Paulo, 13560-970, S\~{a}o Carlos, SP, Brazil.}
\author{I. D'Amico}
\affiliation{S\~{a}o Carlos Institute of Physics, University of S\~{a}o Paulo, 13560-970, S\~{a}o Carlos, SP, Brazil.}
\affiliation{Department of Physics, University of York, York YO10 5DD, United Kingdom.}
\author{V. V. Fran\c{c}a$^*$}
\affiliation{Institute of Chemistry, S\~{a}o Paulo State University, 14800-060, Araraquara, S\~{a}o Paulo, Brazil.}

\begin{abstract}

{\bf Abstract $-$} Metric spaces are characterized by distances between pairs of elements. Systems that are physically similar are expected to present smaller distances (between their densities, wave functions and potentials) than systems that present different physical behaviors. For this reason metric spaces are good candidates {for probing} quantum phase transitions, since they could identify regimes of distinct phases. Here we apply metric space analysis to explore the transitions between the several phases in spin imbalanced systems. In particular we investigate the so-called FFLO (Fulde-Ferrel-Larkin-Ovchinnikov) phase, which is an intriguing phenomenon in which superconductivity and magnetism coexist in the same material. This is expected to appear for example in attractive fermionic systems with spin-imbalanced populations, due to the internal polarization produced by the imbalance. The transition between FFLO phase (superconducting phase) and the normal phase (non-superconducting) and their boundaries have been subject of discussion {in recent} years.  We consider the Hubbard model in the attractive regime for which Density Matrix Renormalization Group calculations allow us to obtain the exact density function of the system. We then analyze the exact density distances as a function of the polarization. We find that our distances {display signatures} of the {distinct quantum phases} in spin-imbalanced fermionic systems: with respect to a central {reference} polarization, systems without FFLO present a very symmetric behavior, while systems with phase transitions are asymmetric. \\

{\bf Keywords $-$} quantum phase transitions; FFLO superconductivity; spin-imbalanced systems; metric spaces.
\\

{*Corresponding author: vvfranca@iq.unesp.br, phone: ++551633019719, fax: ++551633222308 }
\end{abstract}

 \maketitle

\section{Introduction}

Quantum phase transitions are characterized by sudden changes in the physical properties of the system driven by one of its {parameters}. The transition {may be of} first order, {which has a discontinuity in the first derivative with respect to the parameter, or it may be smother}, as in second order transitions and crossovers \cite{sachdev}. Ideally to investigate a quantum phase transition one needs to obtain the order parameter, but this is not trivial in most of the cases. Therefore it is quite common to use other properties of the system {for example witnesses} of quantum phase transitions, such as quantum correlations and entanglement measures \cite{fazio, amico, physica17, canovi, stas, sarandy}.

In this context metric-spaces {analysis appears} as a powerful mathematical tool to investigate {these} transitions. In metric space one can assign distances between wave functions, densities and external potentials of two systems, which quantify the closeness between the systems. Recently natural distances have been proposed for physical systems \cite{prl2011, reply, sharp} and applied in several contexts \cite{sci18, sharp2, sharp3, nagy1, epl15, nagy2, skelt, marocchi}. 

In spin-imbalanced fermionic systems several phases are expected to emerge across the imbalance strength. In the regime of attractive interactions, in particular, the system may present a conventional superfluid phase, as described by the Bardeen-Cooper-Schrieffer theory (BCS phase) \cite{bcs}, a normal magnetic (non-superfluid) phase, a fully polarized (FP) phase and a fascinating phenomenon denominated Fulde-Ferrell-Larkin-Ovchinnikov (FFLO) phase \cite{1, 2}. The FFLO phase is characterized by the exotic coexistence of superconductivity and magnetism and expected to survive against the normal regime for small imbalances. Experimentally, there is only indirect evidences of the FFLO phase: in solid-state materials\cite{organic} and in one-dimensional {Fermi gases} \cite{17}. 

From the theoretical point of view, regardless of the complexity of handling many-particle interactions and the harmonic confinement necessary to describe state-of-the-art experiments, considerable understanding of the FFLO general properties has been achieved \cite{physica17, 3, kim18, fflo1, 10, batrouni, stab}. However the regime of polarizations at which the FFLO phase can be found is still {under} debate, as well as the nature of the transition to the other phases of the spin-imbalanced system.

Here we apply a metric-space analysis to investigate the transitions between BCS, FFLO, normal and FP phases in spin-imbalanced systems described by the one-dimensional fermionic Hubbard model. We obtain the exact density functions of finite but large chains via Density Matrix Renormalization Group (DMRG) calculations and thus calculate density distances between systems with different polarizations. {We find} that the metric spaces are a suitable tool to detect quantum phase transitions. {Our results suggest {\it i)} that} the BCS-normal and the FFLO-normal phase transitions are of first order {and {\it ii)} that} the normal-FP transition is a second-order phase transition or even a mere crossover.

\section{Theoretical and computational methods}

We consider the one-dimensional homogeneous Hubbard model \cite{hubb}:

\begin{eqnarray}
\hspace{-0.4cm}H=-t\sum_{i,\sigma}\left (\hat c_{i,\sigma}^{\dagger}\hat c_{i+1,\sigma}+\hat c_{i+1,\sigma}^{\dagger}\hat c_{i,\sigma} \right) +U\sum_{i}\hat{n}_{i,\uparrow}\hat{n}_{i,\downarrow},
\label{eqn:HubbardHamiltonian}
\end{eqnarray}
where $t$ is the hopping parameter, $U$ is the intra-site interaction and $\hat c_{i,\sigma}^{\dagger},\hat c_{i,\sigma}$ are creation and annhilation operators of fermionic particles at site $i$ with $z$-spin component $\sigma$, up (+1/2) or down (-1/2). Although it is one of the simplest model to describe itinerant and interacting particles in a chain, the Hubbard model describes very important phenomena \cite{hubb2} and has been proved to model properly nanostructures \cite{nanoa, nano1, nanob, nano2, nano3, nano4} and disordered systems \cite{disorder}.

We focus on finite chains, with size $L=80$, polarization {(or imbalance)} quantified by
\begin{equation}
P=\frac{N_{\uparrow}-N_{\downarrow}}{N},
\end{equation}
fixed number of particles $N=N_{\uparrow}+N_{\downarrow}$ and average particle density {(filling factor)} $n=N/L$. Here the spatial inhomogeneity is only due to finite size effects and the density profile is obtained via DMRG techniques \cite{dmrg}.

In all calculations we will use attractive interactions ($U<0$) in units of $t$ and set $t=1$. At $P=0$ the system is expected to be a conventional BCS superfluid, {while} at $P=1$ a fully polarized magnetic system. {For} intermediate polarizations we should have the FFLO phase {up to a certain critical value, $P_C$}  \cite{fflo1}, and a normal non-superfluid phase for $P>P_C$. The metric-space analysis will be {performed here to distinguish between these several phases}.  

We thus consider the metric for densities as discussed in Ref.~\cite{prl2011}:

\begin{eqnarray}
 D(\rho_{1},\rho_{2})=\frac{1}{2N}\int\abs{\rho_{1}(\mbf{r})-\rho_{2}(\mbf{r})} d\mbf{r},
\end{eqnarray}
  where the scaling factor $2N$ was added such that {distances are within} $[0,1]$ to facilitate straightforward comparisons. Here $\rho_{1}$ and $\rho_2$ are the {particle} density functions of any two systems: the more {dissimilar} the systems, the greater the distance.
  
In order to identify the several phases that emerge {by sweeping} the polarization $P$, we explore the distance between {densities corresponding} to different $P$'s. We {use} the central polarization value, $P=0.5$, {to define} the reference system, $\rho_{ref}\equiv \rho_{P=0.5}$ and thus quantify the density distance between this central reference and the system at $P$:
  \begin{eqnarray}
 D(P)=\frac{1}{2N}\int\abs{\rho_P(\mbf{r})-\rho_{ref}(\mbf{r})} d\mbf{r}. \label{drho}
\end{eqnarray}

The choice of the reference system is very important, as it defines which similarities and differences of the systems will be revealed by the distance measure.  We have chosen {it} such that it does not correspond to any of the $P$'s related to the {transitions between} distinct phases (neither $P=0$, $P=P_C$ nor $P=1$) and we have also avoided a reference system within the FFLO regime ($0<P<P_C$, where $P_C^{max}=1/3$ as shown in Ref.~\cite{fflo1}). This particular choice: {\it i)} avoids to emphasize characteristics of only one of the {phase transitions}, {\it ii)} avoids possible strong fluctuations due to the inherent inhomogeneity of the FFLO phase, and {\it iii)} {allows us} to explore a possible symmetry: in the absence of any phase transition, {i.e. if the system presents only the normal phase within $0<P<1$, there should exist a symmetric behaviour} with respect to the central point $P=0.5$, once the differences would be  triggered exclusively by $P$. 
  
\section{Results and discussion}

\begin{figure}[!h]
\centering
  \includegraphics[width=.5\textwidth]{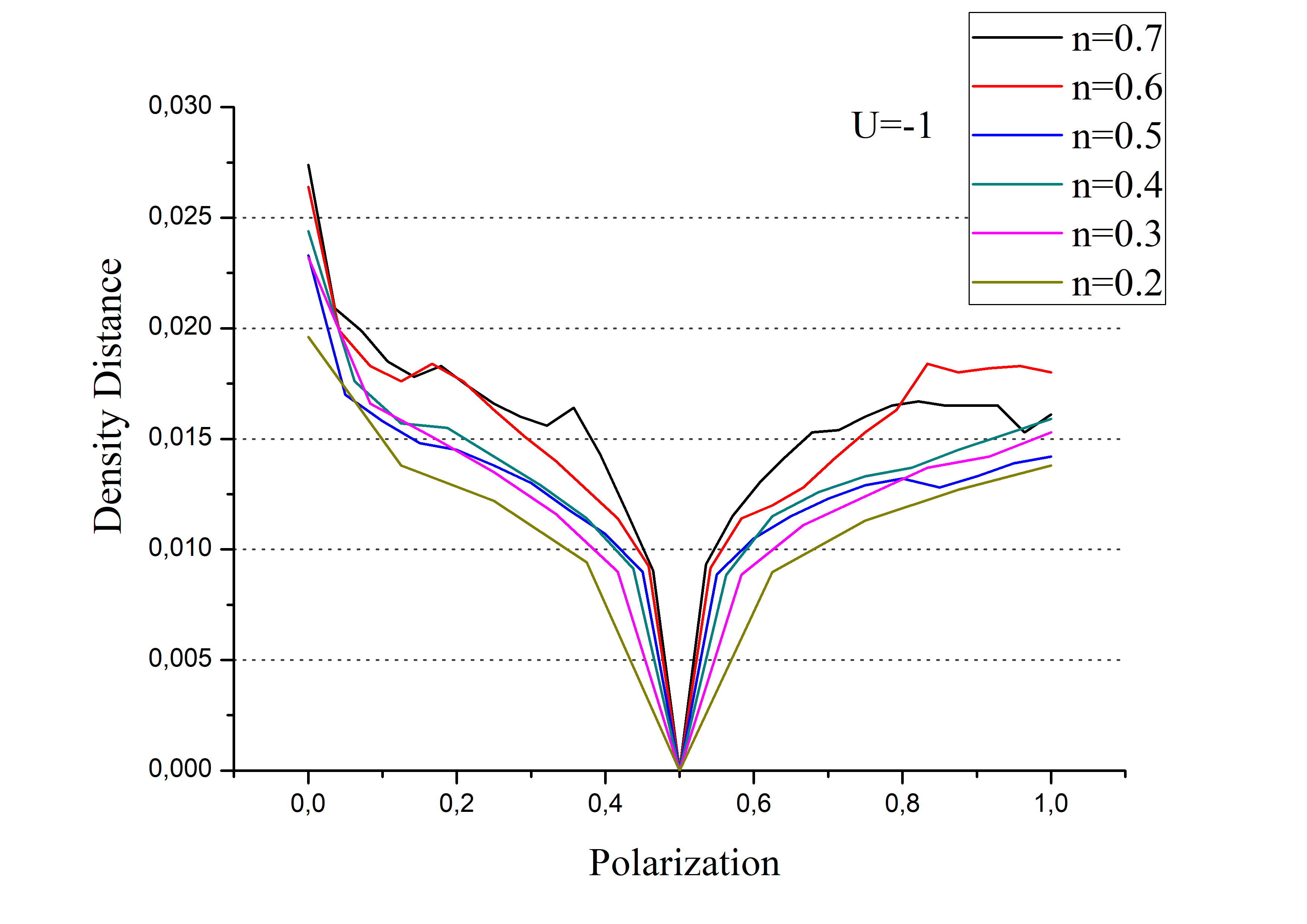}
  \includegraphics[width=.5\textwidth]{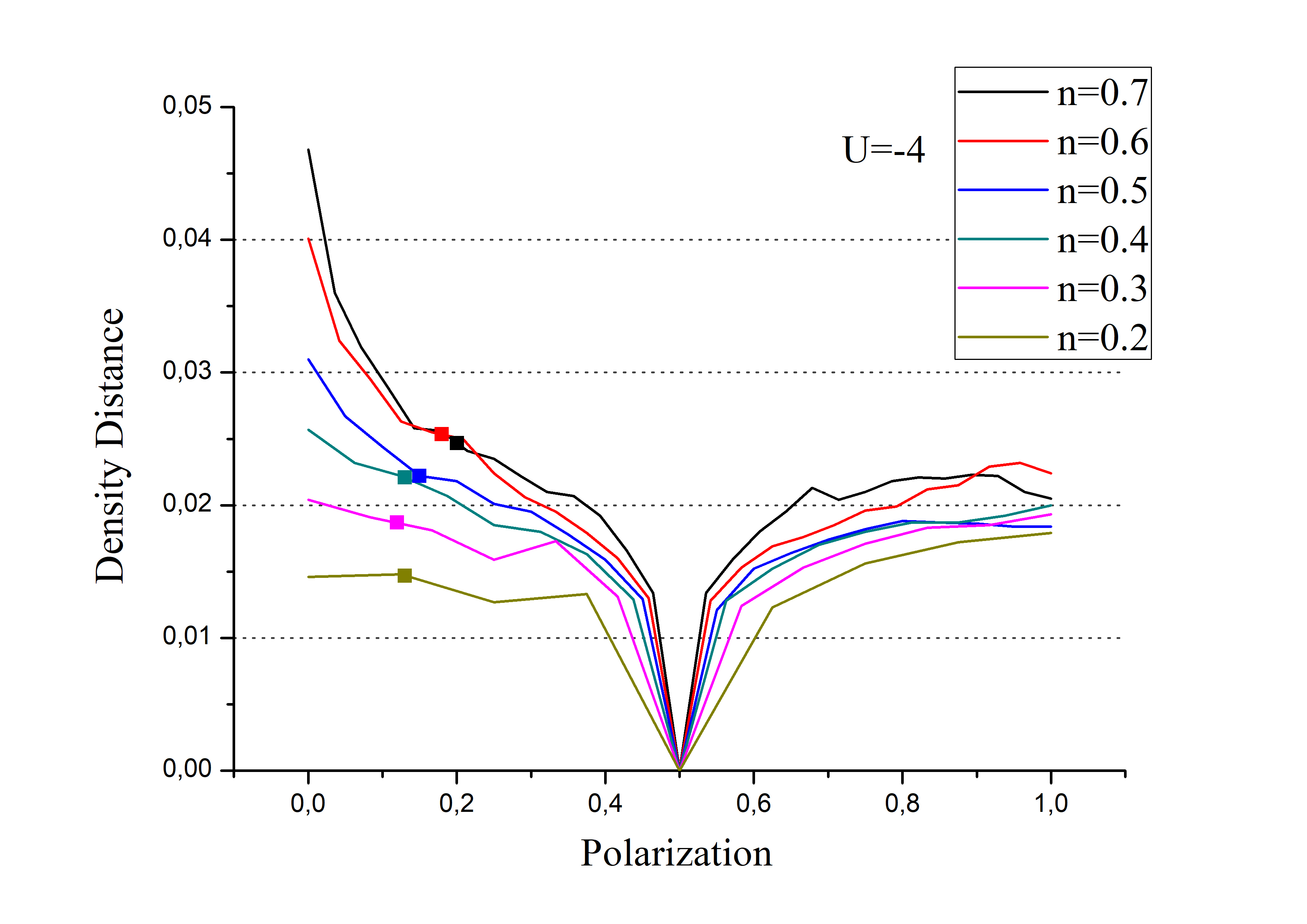}
   \includegraphics[width=.5\textwidth]{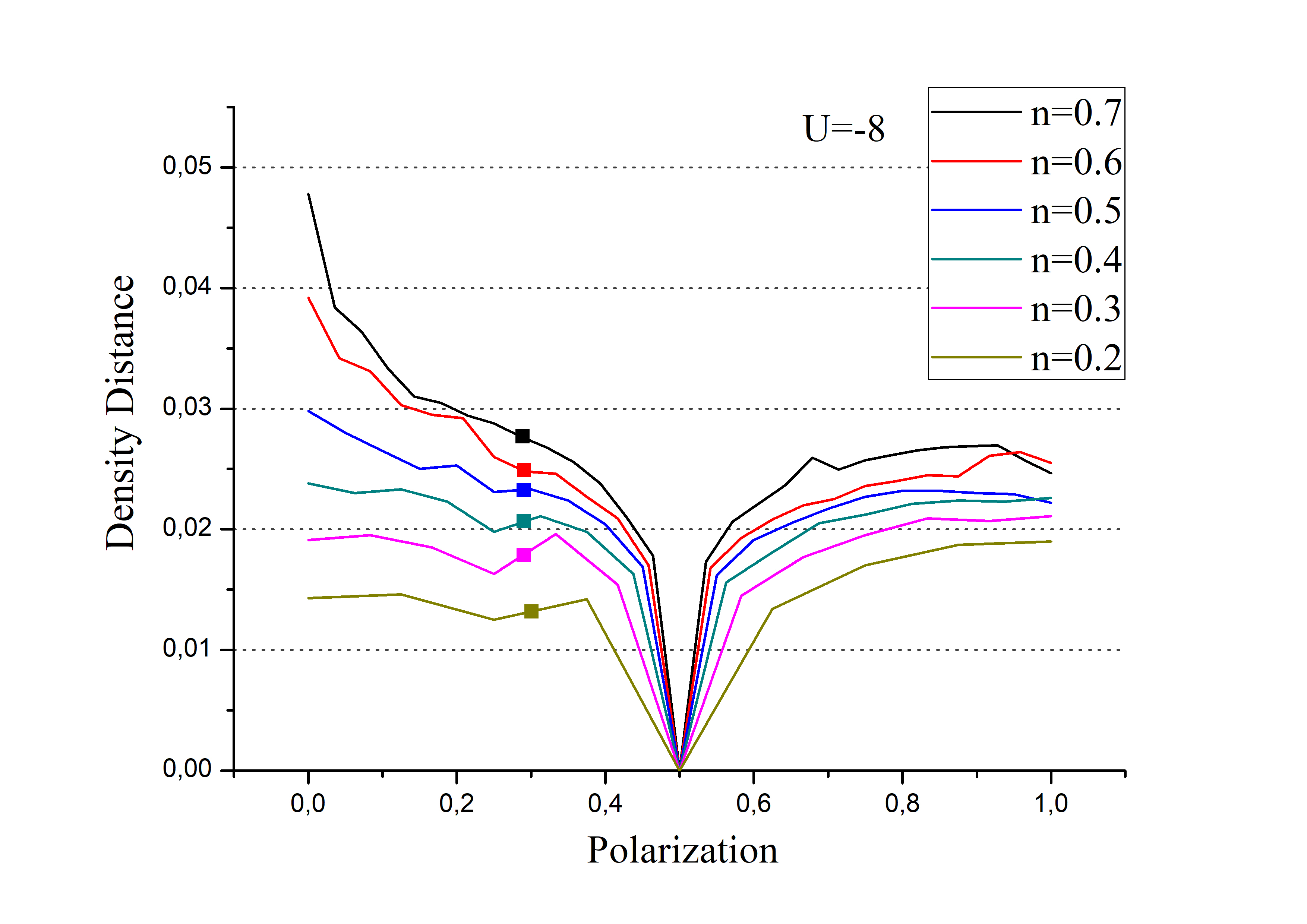}
  \caption{Density distances $D$ as a function of $P$ for several average densities: {upper panel shows non-FFLO systems, while middle and bottom panels present FFLO systems, where solid squares indicate the related $P_C$.} }\label{fig:non_fflo}
\end{figure}

We start by applying our density distance (Eq.~\ref{drho}) in systems without FFLO phase. This may occur when the attractive interaction is too small (here we use $U=-1$), such that the systems undergoes directly {a transition} from the BCS superfluid to the normal phase. So our systems without FFLO are {characterized} by three phases: BCS at $P=0$, normal non-superfluid phase for $0< P< 1$ and fully polarized magnetic phase at $P=1$. 
 
\begin{figure}[!h]
\centering
 \includegraphics[width=.5\textwidth]{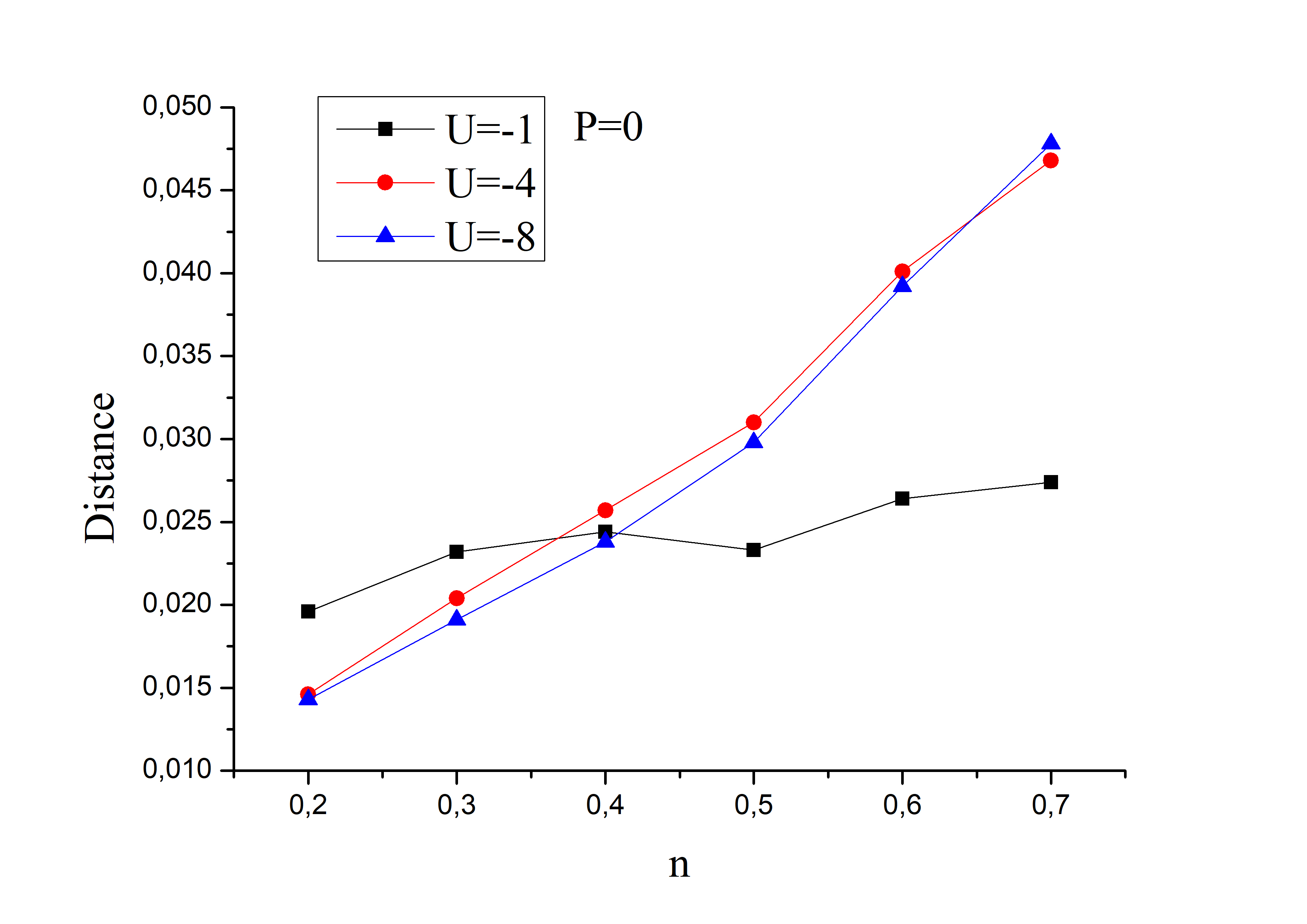}
  \includegraphics[width=.5\textwidth]{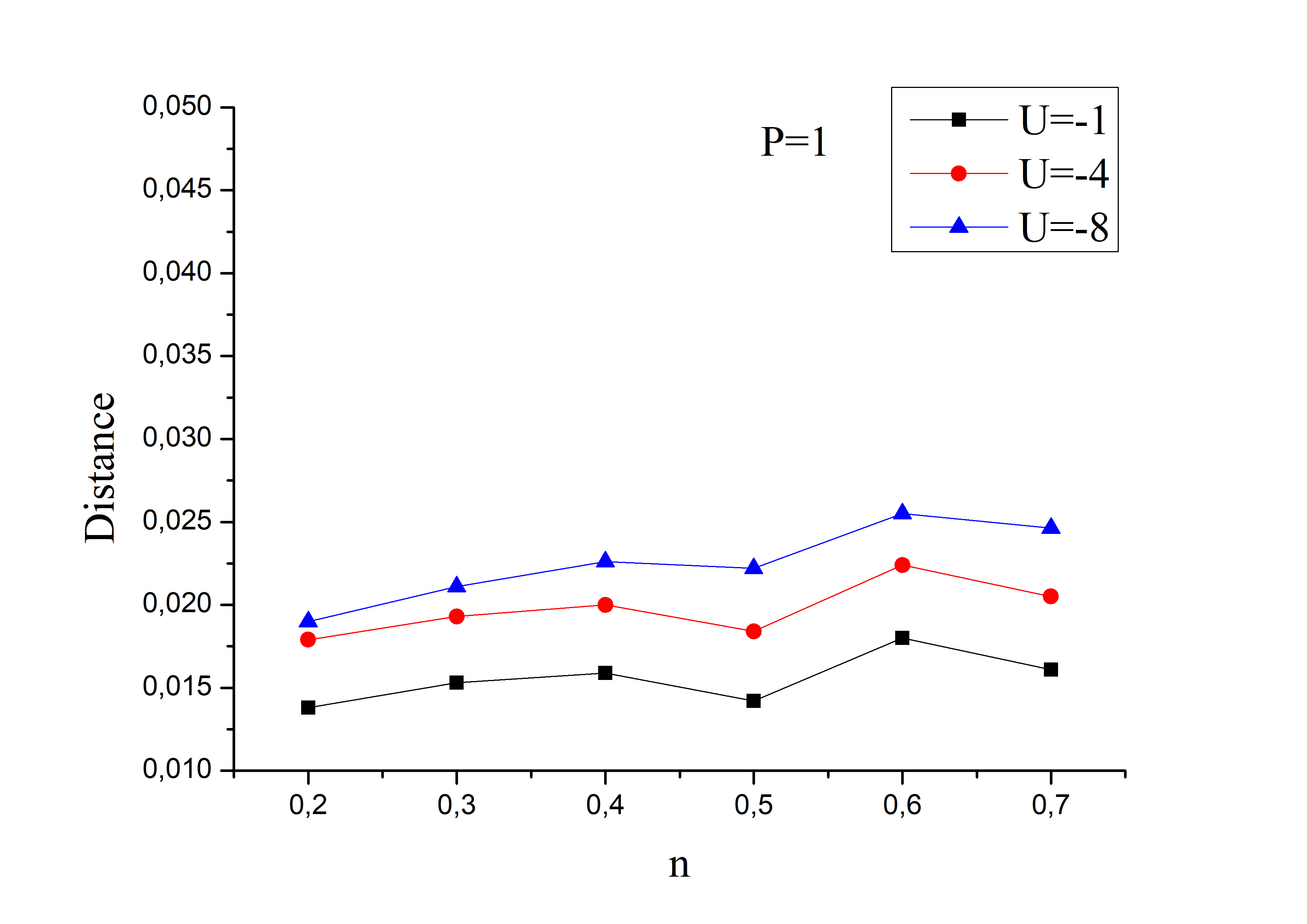}
\caption{{Density distances as a function of the average density (filling factor) for the BCS phase (upper panel) and the FP phase (bottom panel).} }\label{fig:dist_n}
\end{figure}

 In the upper panel of Figure \ref{fig:non_fflo} we present the density distance as a function of $P$ for systems without FFLO phase. {The distance} is essentially symmetric with respect to the central point {for $0.1\lesssim P\lesssim 0.9$}, confirming the fact that there is no phase transition within $0<P<1$. In contrast though there is a clear asymmetry between the BCS and the FP cases. {This can be seen quantitatively in Figure \ref{fig:dist_n}:} the distance at the BCS phase ($P=0$) is considerably larger than at the FP phase ($P=1$). That means that the FP and the normal phase (the {reference}) are physically closer than the BCS and the normal phases. This is not only reasonable $-$ that magnetic phases (normal and FP) are closer than magnetic and superfluid ones (normal and BCS), but it is also consistent with previous works \cite{3, t1, t2}, which suggest that the BCS-normal transition is a first order phase transition, while the normal-FP is a second order phase transition. Thus our distance properly reveals all the features of the system without FFLO: greater distance at $P=0$, consistent with a first order transition, for {$0.1\lesssim P\lesssim0.9$} an almost symmetric behavior with respect to $P=0.5$, confirming the absence of any other transition and, finally the suggestion that {from the normal to the FP the transition} is possibly of second order or simply a crossover.  
 
\begin{figure}[!h]
\centering
 \includegraphics[width=.5\textwidth]{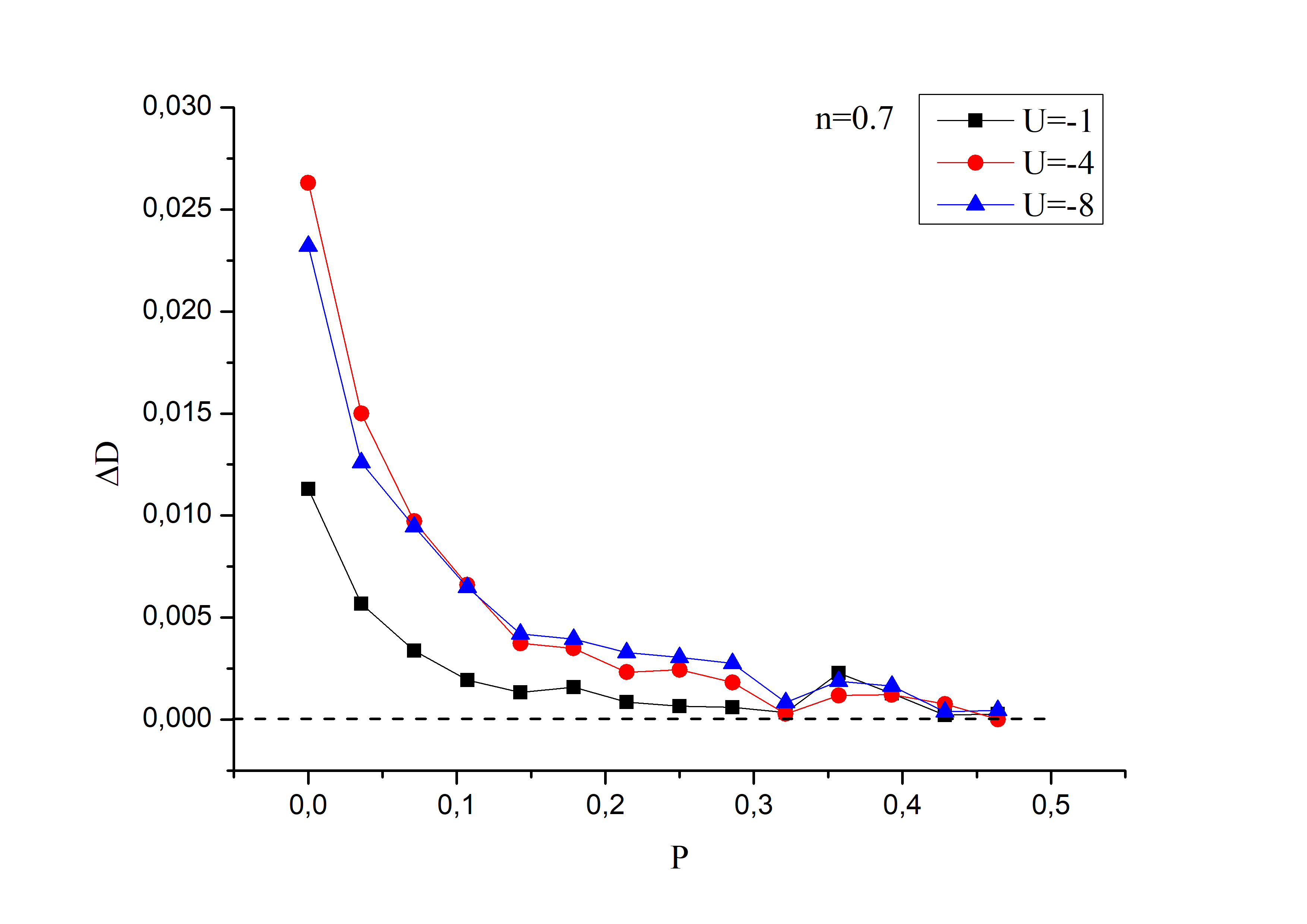}
  \includegraphics[width=.5\textwidth]{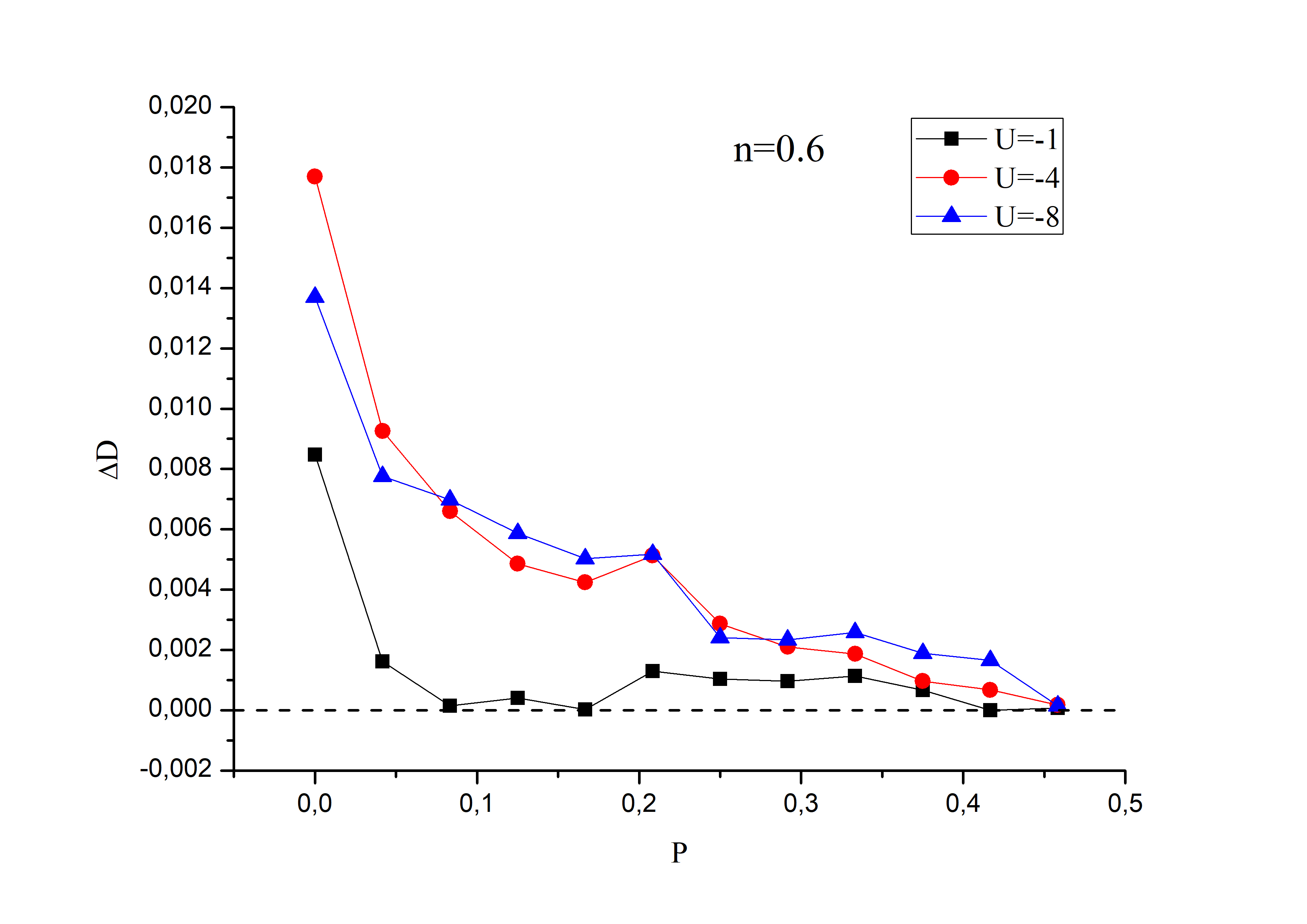}
   \includegraphics[width=.5\textwidth]{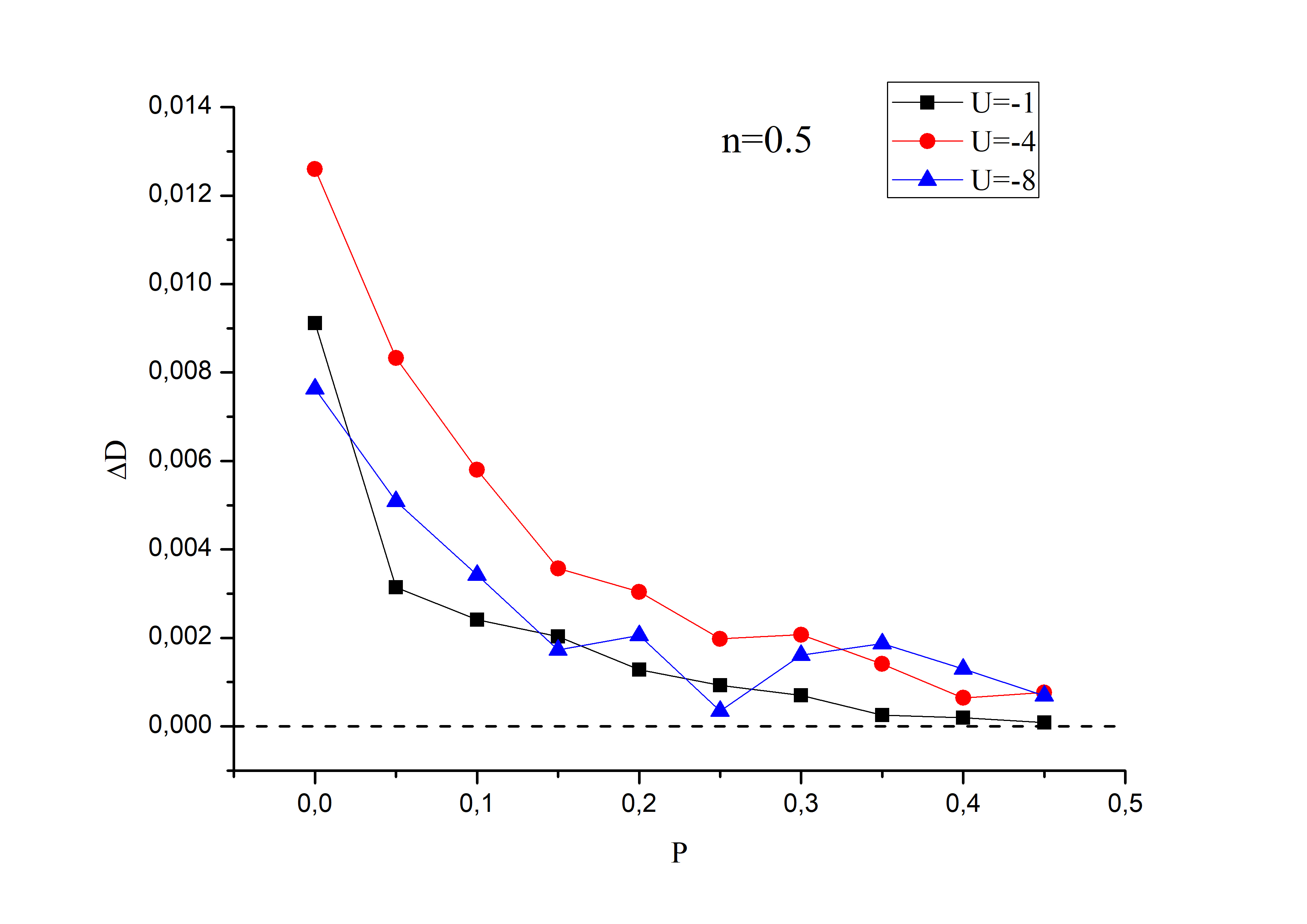}
\caption{Difference between equidistant density distances $\Delta D$, with respect to the central $P=0.5$, as a function of $P$ for non-FFLO systems ($U=-1$) and FFLO systems ($U=-4$ and $U=-8$), for distinct average densities: $n=0.7$ (upper panel), $n=0.6$ (middle panel) and $n=0.5$ (lower panel). }\label{fig:diff}
\end{figure}
 
Now we consider systems with FFLO, which present, in addition to the BCS and the FP phases, the exotic superfluid phase for $0<P\leq P_C$ and the normal magnetic phase for $P_C<P<1$. The intermediate and the bottom panels of Figure \ref{fig:non_fflo} present the results for the density distance in these systems. In contrast to the previous non-FFLO cases, we now see that the asymmetry appears not only at {$P\sim0$}, but also for $P>0$ (except for {$n\lesssim0.4$}). This proves that our distance measure is sensing the FFLO-normal phase transition, and strongly suggests that the FFLO-normal phase transition is also of first order, as the BCS-normal one. For smaller densities,  {$n\lesssim0.4$}, none of the asymmetries appear, what we attribute to the fact that with small number of particles we don't have enough data in such finite systems. Therefore our distance reveals all the {transitions} of the system with or without FFLO:  {we find} larger distances at $P=0$ and for $P<P_C$, suggesting that both BCS-normal and FFLO-normal phase transitions are of first order, while we find no special behavior at $P=1$, suggesting that the normal-FP transition is a crossover or of second order.

Aiming a deeper understanding of the asymmetric behavior we explore the following difference between distances:
{
\begin{equation}
\Delta D (P)= |D(1-P)-D(P)|,
\end{equation}}
which quantifies the asymmetry between two equidistant polarizations with respect to the central, $P=0.5$. As one can see in Figure \ref{fig:diff}, in general $\Delta D$ is larger for the systems with FFLO ($U=-4$ and $U=-8$) than for systems without FFLO ($U=-1$). Despite the fact that there are some fluctuations, it is clear that for non-FFLO systems $\Delta D$ is significant only at the BCS-normal transition, while for FFLO systems it is significant also for $P>0$. However $\Delta D$ is not able to identify precisely the $P_C$, where the FFLO-normal phase transition appears. We attribute the difficulty to precisely determine $P_C$ to the intrinsic density fluctuations of the inhomogeneous superfluid FFLO phase: {this contributes additional fluctuations to} the distances and therefore a fair comparison between densities within the FFLO regime becomes impracticable.

 \section{Conclusions}

In summary, {we used metric-spaces analysis} to probe quantum phase transitions. We find that our distances are able to distinguish among the quantum phases present in spin-imbalanced fermionic systems. Our results $-$ via a distance measure {defined} with respect to a {reference} at the normal phase regime $-$ {suggest} that both the BCS-normal and the FFLO-normal phase transitions are of first order, while the normal-FP phase transition is a smoother transition, of second order or simply a crossover. Future work include {refining} the metrics such that one can identify precisely the {critical $P$ values} and to apply the metric-spaces approach for investigating other quantum phase transitions.

\acknowledgements

VVF is financially supported by FAPESP (Grant: 2013/15982-3) and CNPq (Grant: 448220/2014-8). IDA acknowledges support from CNPq (Grant: PVE - Processo: 401414/2014-0) and from the Royal Society through the Newton Advanced Fellowship scheme (Grant no. NA140436).




\begin{thebibliography}{99}
\bibitem{sachdev} S. Sachdev, {\it Quantum Phase Transitions}, Cambridge University Press (2011).
\bibitem{fazio} A. Osterloh, L. Amico, G. Falci, R. Fazio, Nature {\bf 416}, 608 (2002).
\bibitem{amico} L. Amico, R. Fazio, A. Osterloh, V. Vedral, Rev. Mod. Phys., {\bf 80}  517 (2008).
\bibitem{physica17} V. V. Fran\c{c}a, Physica A {\bf475}, 82 (2017).
\bibitem{canovi} E. Canovi, E. Ercolessi, P. Naldesi, L. Taddia, D. Vodola, Phys. Rev. B {\bf 89}, 104303 (2014).
\bibitem{stas} J. Stasinska, B. Rogers, M. Paternostro, G. De Chiara, A. Sanpera, Phys. Rev. A {\bf 89}, 032330 (2014).
\bibitem{sarandy} L.-A. Wu, M. S. Sarandy, D. A. Lidar, and L. J. Sham, Phys. Rev. A {\bf74}, 052335 (2006).
\bibitem{prl2011} I. D'Amico, J. P. Coe, V. V. Fran\c{c}a, and K. Capelle,  Phys. Rev. Lett. {\bf106}, 050401 (2011).
\bibitem{reply} I. D'Amico, J. P. Coe, V. V. Fran\c{c}a, and K. Capelle Phys. Rev. Lett. {\bf107}, 188902 (2011).
\bibitem{sharp} P. M. Sharp and I. D'Amico, Phys. Rev. B {\bf89}, 115137 (2014).
\bibitem{sci18} V. V. Fran\c{c}a, J. P. Coe, and I. D' Amico, Sci. Rep. {\bf8}, 664 (2018).
\bibitem{sharp2} P. M. Sharp and I. D'Amico, Phys. Rev. A {\bf92}, 032509 (2015).
\bibitem{sharp3} P. M. Sharp and I. D'Amico,
Phys. Rev. A {\bf94}, 062509 (2016).
\bibitem{nagy1} I. Nagy and I. Aldazabal
Phys. Rev. A {\bf84}, 032516 (2011).
\bibitem{epl15} J. P. Coe, V. V. Fran\c{c}a and I. D'Amico, EPL {\bf110}, 63001
(2015).
\bibitem{nagy2} \'A. Nagy and E. Romera
Phys. Rev. A {\bf88}, 042515 (2013).
\bibitem{skelt} A. H. Skelt, R. W. Godby, I. D'Amico, arXiv:1801.04132 (2018).
\bibitem{marocchi} Simone Marocchi, Stefano Pittalis, and Irene D'Amico
Phys. Rev. Materials {\bf1}, 043801 (2017).
\bibitem{bcs} J. Bardeen, L. N. Cooper, J. R. Schrieffer, Phys. Rev. {\bf106}, 162 (1957).
\bibitem{1} P. Fulde, R. A. Ferrell, Phys. Rev. {\bf 135}, A550 (1964). 
\bibitem{2} A. L. Larkin, Y. N. Ovchinnikov, Sov. Phys. JETP {\bf20}, 762 (1965).
\bibitem{organic} H. Mayaffre, S. Krä\"amer, M. Horvati\'c, C. Berthier, K. Miyagawa, K. Kanoda, V. F. Mitrovi\'c, Nature Physics {\bf10}, 928 (2014). 
\bibitem{17} Y. Liao, A. S. C. Rittner, T. Paprotta, W. Li, G. B. Partridge, R. G. Hulet, S. K. Baur, E. J. Mueller, Nature {\bf467}, 567 (2010).
\bibitem{3} R. Casalbuoni, G. Nardulli, Rev. Mod. Phys. {\bf76}, 263 (2004).
\bibitem{kim18} Jami J Kinnunen, Jildou Baarsma, Jani-Petri Martikainen and Paivi Torma, Rep. Prog. Phys. in press (2018).
\bibitem{fflo1} V. V. Fran\c{c}a, D. H\"{o}rndlein, A. Buchleitner, Phys. Rev. A {\bf 86}, 033622 (2012).
\bibitem{10} D. -H. Kim, J. J. Kinnunen, J. -P. Martikainen, P. T\"orm\"a, Phys. Rev. Lett. {\bf106}, 095301 (2011).
\bibitem{batrouni} G. G. Batrouni, M. H. Huntley, V. G. Rousseau, R. T Scalettar, Phys. Rev. Lett. {\bf 100}, 116405 (2008).
\bibitem{stab} H. Caldas, M. A. Continentino, J. Phys. B: At. Mol. Opt. Phys. {\bf 46}, 155301 (2013).
\bibitem{hubb} J. Hubbard, Proc. R. Soc London. Series A {\bf276}, 238 (1963).
\bibitem{hubb2} Klaus Capelle, Vivaldo L. Campo Jr., Phys. Rep. {\bf528}, 91 (2013).
{\bibitem{nanoa} J. Fern{\'a}ndez-Rossier, J\_J Palacios, Phys. Rev. Lett. {\bf 99}, 177204 (2007).}
\bibitem{nano1}J. P. Coe, V. V. Fran\c{c}a, and I. D'Amico, Phys. Rev. A {\bf81},
052321 (2010).
{\bibitem{nanob}O. V. Yazyev, Rep. Prog. Phys. {\bf 73}, 056501 (2010).}
\bibitem{nano2}J. P. Coe, V. V. Fran\c{c}a, and I. D'Amico, EPL {\bf93}, 10001 (2011).
\bibitem{nano3}J. P. Coe,  V. V.  Fran\c{c}a, and I. D'Amico, J. Phys. Conf. Series {\bf 286}, 012048 (2011).
\bibitem{nano4} T. Mendes-Santos, T. Paiva, R. R. dos Santos, Phys. Rev. B {\bf 87}, 214407 (2013).
\bibitem{disorder} V. V. Fran\c{c}a, I. D' Amico, Phys. Rev. A {\bf 83}, 042311 (2011).
\bibitem{dmrg} U. Schollw\"ock, Rev. Mod. Phys. {\bf77}, 259 (2005).
{
\bibitem{t1} S. Pilati, S. Giorgini, Phys. Rev. Lett. {\bf 100}, 030401 (2008).
\bibitem{t2} D. E. Sheehy, L. Radzihovsky, Ann. of Phys {\bf 322}, 1790 (2007).}












\end{thebibliography}
\end{document}